\documentclass[conference]{IEEEtran}
\ifCLASSINFOpdf
\else
\fi

\usepackage{wspr}
\let\labelindent\relax
\usepackage{enumitem}
\usepackage{amsmath,amssymb,amsfonts}
\usepackage{algorithmic}
\usepackage{makecell}
\usepackage{booktabs}
\usepackage{graphicx}
\usepackage{textcomp}
\usepackage{xcolor}
\usepackage{float}
\usepackage{cite}
\usepackage{url}
\def\BibTeX{{\rm B\kern-.05em{\sc i\kern-.025em b}\kern-.08em
    T\kern-.1667em\lower.7ex\hbox{E}\kern-.125emX}}

\newcounter{rqcounter}
\begin{document}
\title{A First Look at Scams on YouTube}

\author{\IEEEauthorblockN{Elijah Bouma-Sims}
\IEEEauthorblockA{North Carolina State University\\
erboumas@ncsu.edu}
\and
\IEEEauthorblockN{Bradley Reaves}
\IEEEauthorblockA{North Carolina State University\\
bgreaves@ncsu.edu}}

\IEEEoverridecommandlockouts
\makeatletter\def\@IEEEpubidpullup{6.5\baselineskip}\makeatother
\IEEEpubid{\parbox{\columnwidth}{
    Workshop on Measurements, Attacks, and Defenses for the Web 2021\\
    February 25, 2021\\
    ISBN 1-891562-66-5\\
    https://dx.doi.org/10.14722/ndss.2021.23001\\
    www.madweb.work
}
\hspace{\columnsep}\makebox[\columnwidth]{}}

\maketitle

\begin{abstract}
YouTube has become the second most popular website according to Alexa, and
it represents an enticing platform for scammers to attract victims.
Because of the computational difficulty of classifying multimedia, identifying scams on YouTube is  more difficult than text-based media. As a consequence, the research community to-date has provided little insight into the prevalence, lifetime, and operational patterns of scammers on YouTube.
In this short paper, we present a preliminary exploration of scam videos on
YouTube.
We begin by identifying 74 search queries likely to lead to scam videos
based on the authors' experience seeing scams during routine browsing. We then
manually review and characterize the results to identify 668 scams in 3,700 videos.
In a detailed analysis of our classifications and metadata, we find that
these scam videos have a median lifetime of nearly nine months, and many rely on
external websites for monetization. We also explore the potential of
detecting scams from metadata alone, finding that
metadata does not have enough predictive power to  distinguish scams  from legitimate videos.
Our work demonstrates that scams are a real problem for YouTube users,
motivating future work on this topic.
\end{abstract}

\begin{comment}
  Fraud is a persistent problem online.
  In particular, the popular video host YouTube is home to many scam videos which attempt to defraud users.  
  Despite the threat they pose to the website’s large
  user base, prior work has not characterized the problem. We
  use YouTube’s search API and related keywords, such as
  ”free Amazon gift card” or ”Facebook.com tech support,”
  and identify fraud videos among the top results.  Out of 3700  
  search results, we have identified at least 668 videos which 
  are considered scams based on YouTube's content policy
  or United State's law and 467 of which remained after 6 months. We characterize the features of these videos and their user created meta data, finding differences in age and viewership.  The types of profiles which post these videos are also analyzed, revealing that profiles hosting scams are generally inactive. Finally, we discuss possible methods of addressing this issue based on our results. Ultimately, our results will provide for more directed work targeting scam videos as well as inform new policies to reduce the prevalence of video fraud on the website. They also show that scam videos are a real issue on YouTube, particularly among search terms related to mobile game currency and gift cards.

\end{comment}

%\begin{IEEEkeywords}
%YouTube, Consumer protection, Metadata
%\end{IEEEkeywords}

\section{Introduction}

As the most popular video sharing and streaming platform and the 
second most popular website on the Internet according to Alexa~\cite{noauthor_alexa_nodate}, YouTube is a 
natural target for the perpetration of online fraud
and other computer crimes.
%~\cite{ali_consumer} 
%~\cite{zhang_survey}  
While there is a large body of work looking at other misuse on the platform such as
harassment~\cite{aggarwal_metadata}, extremism~\cite{sureka_extremists,conway_jihad} ,
video spam~\cite{Benevenuto_video_spam}, hate~\cite{agarwal_extremism} ,
disinformation~\cite{hussain_disinformation_youtube,mohammed_conspiracy},  and 
more~\cite{Papadamou_disturbed}, to the best of our knowledge no paper has examined how
scammers use video sharing sites. 

To this end, we present a preliminary analysis of scam videos on YouTube.
We begin by collecting 3,700 YouTube videos from search queries likely to lead to scams based on authors' experiences on the platform.
We then manually analyze these videos to determine if they meet our criteria for a scam video; we derived these criteria from YouTube's content policy and Terms of Service and United States law. 
We determined 668 videos met these criteria, with examples including a video that pretends to provide the user with 
``free'' Walmart gift cards by ``hacking'' Walmart, 
a video that pretends to ``generate'' free Fortnite currency,
and a video that claims to provide free Door Dash credits to viewers.
We then collect metadata about the video, including user interactions and lifetime, as well as captions when available. 
We also collect metadata on the YouTube channel of the scam video owner. Finally, we revisit these videos five months later to
determine liveness.

In this paper, we use the collected dataset to address the following research questions:

\begin{enumerate}[label=\textbf{RQ\arabic*}]
    \item \textit{How does the metadata differ between scam and non-scam videos?} Scam videos are younger than non-scam videos, have fewer views and less comment engagement, and are posted by accounts with less activity than non-scam videos.
    \item \textit{How do scammers monetize scams on YouTube?} The vast majority of scammers redirected to external websites, many of which use ``Cost per action'' (CPA) monetization tools like surveys to make money. Further, most scams in our dataset were on search terms related to gift cards or mobile games. 
    \item \textit{Can a classifier use metadata alone to distinguish scam and non-scam videos?} Statistical fields related to channel size and popularity provided the most mutual information. Additionally, the presence of gift card and mobile game-related words in metadata was found to be significant in discriminating between videos.
\end{enumerate}

The remainder of this paper proceeds as follows.
Section~\ref{sec:data_collection} describes the methodology used to answer our research questions and our data collection system. Section~\ref{sec:results} describes our data and the results of our analysis. Section~\ref{sec:case_studies} describes several important case studies observed in the process of completing this study. Section~\ref{sec:conclusion} provides discussion and future work.
\section{Methods}
\label{sec:data_collection}

\begin{figure}[!tbph]
    \label{fig:giftgenerator}
    \includegraphics[width=\columnwidth]{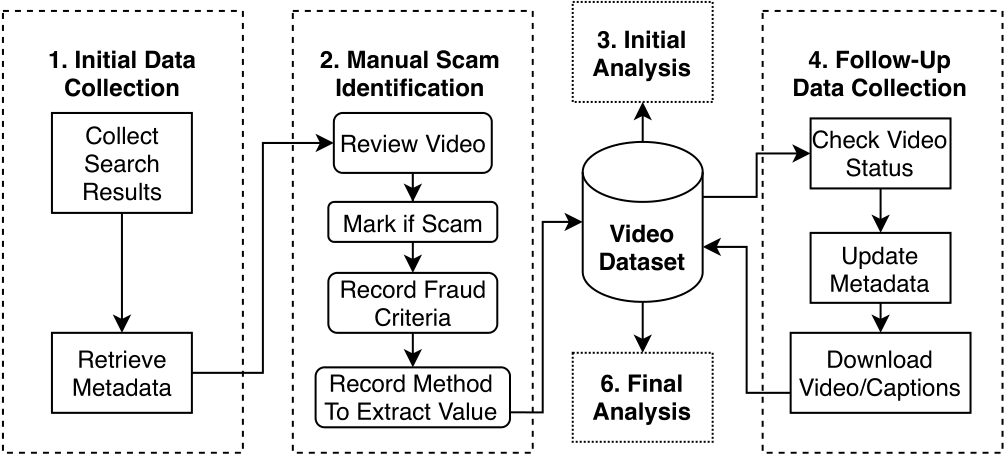}
    \caption{Overview of Analysis Process}
    \label{fig:analysis_flow}
\end{figure}

We begin our methods discussion by describing our definition of a scam video before discussing data collection and initial analysis, then finally describing our follow-up data collection and analysis. We also discuss the limitations of our methodology. An overview of our analysis process can be seen in Figure~\ref{fig:analysis_flow} 

\subsection{Definition of Scam Video}

Before we collected or analyzed data, inspired by Pouryousefi et al.~\cite{Pouryousefi2019} we defined a video scam as \emph{a video that attempts to attract viewers through misrepresentation, including fraudulently offering tangible, intangible, or financial awards.} Since a video view by itself may provide ad revenue, it is not necessary that there be an 
external profit source in our definition of a scam; however, we expected that many scams would attempt to direct users off-site to extract revenue.

To operationalize this definition of scam videos, we
conservatively identify criteria based strictly on United States law and YouTube’s
Terms of Service. If a video falls into any of the following criteria,
it was considered a scam for our purposes:

\begin{itemize}
    \item Content which purports to commit a crime on behalf of the user (whether it actually does or not)
    \item Content which purports to provide an ``unbounded'' giveaway that offers unlimited free items without rules, limit or end
    \item Content which violates the following points from YouTube's content policy on ``Spam, misleading metadata, and scams''~\cite{youtube_spam_policy}:
        \begin{itemize}
            \item{``Promises viewers they'll see something but instead directs them off-site.''}
            \item{``Gets clicks, views, or traffic off YouTube by promising viewers that they’ll make money fast.''}
            \item{``Sends audiences to sites that spread malware, try to gather personal information or other sites that have a negative impact.''}
            \item{``Offering cash gifts, `get rich quick' schemes, or pyramid schemes (sending money without a tangible product in a pyramid structure).''}
        \end{itemize}
    \item Content that impersonates another person, company, or organization. 
    \item Any video whose claimed effect is demonstrably invalid. 
\end{itemize}

\subsection{Initial Data Collection}

With a clear definition of scam videos, we then moved to collect data.
The first challenge in creating a dataset was the likely sparsity of scams. With over 500 hours of videos uploaded to YouTube per minute~~\cite{youtube_for_press}, the amount of scam content is likely to only make up a very small fraction of a random sample of videos on the website. Thus, while a web crawler style approach may provide a representative sample, any sample small enough to be viewed in its entirety would contain virtually no offending content. Additionally, a random sample of all videos on the website would not accurately reflect the behavior of users. A viewer using the website normally does not have an equal chance of finding any given video. Rather, they will be directed to videos related to their interests based on search queries and YouTube's recommended video algorithm. 

Rather than collect a random sample of all videos on the website, we collected videos using the ``search.list'' function of the YouTube Data API~\footnote{\url{https://developers.google.com/youtube/v3}}. The search queries were chosen based on patterns that were expected to produce scam videos. 
We had previously found videos under similar queries during normal YouTube browsing.
Table \ref{search_terms} shows a list of search query patterns and the source used to fill the pattern. Generally speaking, the 74 search queries fell into four categories: tech support (48), mobile game currency (8), bank support (8), and gift cards (24). In each search query pattern, the character * substitutes for the company or software name. The first 50 videos were retrieved for each search query. 

\begin{table}[!tbp]
\caption{Search queries used to generate dataset}
\begin{tabular}{cccc}
\toprule
Category  & Pattern & Source & \# of queries 
\\ \midrule
Gift Card & ``Free * Gift Card'' & \makecell{Amazon top \\ 100 gift cards~\cite{amazon_top_100}} & 24          
\\ \midrule
Tech Support & ``* Tech Support'' & \makecell{Top Tech Companies \\ on Forbes \\ Global 2000 List~\cite{investopedia_tech}} & 8 
\\ \midrule
  Bank Support & ``* Support'' &  \makecell{Top 10 Private Banks \\ According \\ to Investopedia~\cite{investopedia_banks}} & 10 
  \\ \midrule
  Tech Support & ``* Tech Support'' & \makecell{Alexa Top \\ 50 Websites \\ (in English)~\cite{noauthor_alexa_nodate}} & 24 \\ \hline
  Mobile Games & ``Free * Currency'' & \makecell{Top Mobile Games \\ (available in the US) \\ by Worldwide Revenue for \\ January 2020} & 8 
  \\ \bottomrule
\end{tabular}
\label{search_terms}
\end{table}

Once a full list of videos was collected---with a total of 3700 videos---the dataset was filled out with information from the ``video.list'' and ``channel.list'' methods of the YouTube API. The final dataset has 21 fields, including video statistics (number of views, likes, dislikes, and comments), channel statistics (number of views, videos, and 
subscribers), video creation date, channel creation date, and user-created metadata
(title, description, tags, and video description).

The final step of initial data collection was to manually categorize the content of each video. As the videos were evaluated, three data points were recorded: a boolean value corresponding
to whether the video is a scam or not, a list of fraud criteria that were violated
by the video, and the method(s) which the video used to extract value from viewers.
Methods to extract value include redirecting to a website, providing a software download,
and providing a phone number. We ultimately identified 668 scams out of the 3700 videos. These two categories of videos were then analyzed and compared using statistical tests to determine if there was a meaningful difference between scam and non-scam videos (RQ1). The additional two variables collected while viewing videos were used to determine the most common methods used to make money (RQ2). The exact methods and results are described in section~\ref{sec:results}.

\subsection{Follow-Up Analysis}

After 5 months, in September 2020, we reviewed the status of videos from our initial crawl using the YouTube API to determine which videos had been removed.  We did not consider these removed videos in secondary data analysis as their captions or videos could not be downloaded. The Python library PyTube\footnote{\url{https://pypi.org/project/pytube/}} was used to download the MP4 file and English caption files for all videos that had not been removed and which had English captions. Some videos were unable to be downloaded despite being still available, including several videos which had been changed to be restricted to channel members since initial data collection. Ultimately, 322 out of the 3032 non-scam videos were removed, while 201 out of the 668 scam videos were removed. Of those 3165 videos remaining, we were able to download 3078 of the videos and get English captions from 2026 of the videos. 

The YouTube Data API was also used to update metadata fields which had changed since the dataset was initially created. Although the vast majority of the videos did not have changes in their metadata fields, this updated metadata was used along with the aforementioned video data to analyze which metadata and statistical fields may provide the most information for video classification (RQ3). The exact methods and results are described in section~\ref{sec:results}.

\subsection{Limitations}

Like all measurement studies, this work does have some limitations. As described above, our dataset is intentionally biased and cannot be said to be representative of the general population of non-scam or scam videos. Our findings are necessarily limited to the search queries used to seed the dataset. Additionally, our method for identifying scam videos was based on subjective and contextual manual analysis. Classification was done by a single person to reduce variance in how these rules were applied. Finally, our searches returned some non-English videos that could not be evaluated. Our data, thus, applies only to content in English.
\section{Results}
\label{sec:results}
We begin by discussing video statistics before moving to look at the type of scams in our dataset and then finally describing our analysis of different variables for classification. 

\subsection{RQ1: How does the metadata differ between scam and non-scam videos?}
\label{ssec:vidstats}

After creating our dataset, our initial goal was to evaluate the properties of both scams and non-scam videos and determine whether such differences were statistically meaningful. To this end, we used a two-sample t-test to compare the median age, the median number of views, the median rating, and the channel statistics of scam and non-scam videos. While we cannot necessarily assume the normality of the data or the underlying distribution, the t-test is resilient to non-normal data~\cite{nonnormal_stats}. These tests are all based on the initial data collection and are summarized in table~\ref{tab:ttest}. For each test, the null hypothesis was that there was no difference between the two datasets with an alpha of 0.05.

This experiment led to the following findings: 

\paperFinding{Non-scam videos had much more activity than the scam videos in our dataset, but like-dislike was not shown to be a reliable indicator of whether a video is scam or not} Both the number of views and number of comments on scam videos were less than those on non-scam videos ($p < 0.001$). While the median like to dislike ratio of scam videos was less than that of the non-scam videos, the t-test did not show a statistically significant difference ($p = 0.21$). This test did not include videos that had the like and dislike function disabled or videos with no dislikes, but the finding is notable since it means that users may not be able to rely on the ratings as an indication of a video's accuracy. 

\paperFinding{Scam videos are newer than non-scam videos but can still have a long life} The median age of non-scam videos was nearly twice the median age of scam videos ($p < 0.001$). Notably, the median age of scam videos was still 268 days, with ages ranging from 0 to 4440 days. This finding does not necessarily mean that scam videos have been posted more recently than non-scam videos. It is possible that older scam videos are more likely to be flagged and removed from YouTube. Additionally, several of the search terms reference video games that were released within the last two years. However, this does show that scam videos are still being posted, even as soon as several days before the data was collected.

\paperFinding{scam hosting channels have less activity than non-scam channels}  The median number of views, videos, and subscribers were significantly lower for channels that hosted scams as compared to channels without scam videos. This finding was expected, as it is likely hard for scam channels to gain popularity on YouTube. This also suggests that channels may be created or used for individual fraud campaigns before being abandoned. During analysis, we noticed some evidence of channels being taken over for the purposes of disseminating scams, such as channels used for home videos uploading scam videos after years of inactivity. 

\begin{table}[!tbp]
\caption{Comparison of Scam and Non-scam Video Statistics}
\begin{tabular}{cccc}
\toprule
Statistic & \makecell{Median \\ Scam Value} & \makecell{Median \\ Non-Scam Value} & P-Value
\\ \midrule
Age in Days & 268 & 507 & $1.2*10^{-10}$
\\ \midrule
\# of Views & 2519 & 5816 & $4.3*10^{-7}$
\\ \midrule
Like-Dislike Ratio & 8.5 & 20.6 & $0.21$
\\ \midrule
\# of Comments & 8.0 & 23.5 & $2.1*10^{-4}$
\\ \midrule
\# of Channel Videos & 9 & 339 & $2.2*10^{-29}$
\\ \midrule
\# of Channel Views & 16112.5 & 3990403.0 & $3.3*10^{-16}$
\\ \midrule
\# of Channel Subscribers & 35 & 20400 & $9.8*10^{-33}$
\\ \bottomrule
\end{tabular}
\label{tab:ttest}
\end{table}

\subsection{RQ2: How do scammers monetize scams on YouTube?}

As mentioned in the previous section, as the videos in the dataset were viewed, scams were classified based on the method they used to extract value and the rule they violated. We also recorded the search term they were found using. These stats lead to the next three findings: 

\paperFinding{Most scams in the dataset redirected users to other websites to extract value} Out of the 668 scams, 526 (78.7\%) directed viewers to visit an external website. It is possible that they led to an app download or phone number; however, the immediate point of redirection was a web address. 45 (6.74\%) videos directed viewers immediately to download an application, and 44 (6.59\%) videos directed users to call a phone number. 

\paperFinding{Most scams in the dataset were found with search queries related to mobile game currency or gift cards} Of the 668 scams, 393 (58.8\%) were found with searches relating to gift cards, and 218 (32.6\%) were found with searches relating to mobile games. The latter case is even more significant than its magnitude suggests, considering that only 8 search queries were related to mobile games.\textit{ All of those 8 search queries contained scams}. There were 57 (8.53\%) scams on tech support search terms and no scams on bank support-related search terms. Table~\ref{table:Searchterms} shows more detail, providing the number of scams for the eleven search results with at least as many scam videos as non-scam videos in the top 50.

\begin{table}[h]\centering
	\caption{\# of search terms with at least as many scam as nonscams}\centering
		\begin{tabular}{ccc}
			\toprule 
			 Rank & Search Term & \# of Scams 
			\\ \midrule
			1 & 'Free Gardenscapes Coins' & 42 
			\\ \midrule
			2 & 'Free PUBG Battle Points' & 38
			\\ \midrule
			3 & 'Free Candy Crush Gold Bars' & 34 
			\\ \midrule
			4 & 'Free Mastercard Gift Card' & 33 
			\\ \midrule
			5 & 'Free Netflix Gift Card' & 32 
			\\ \midrule
			6 & 'Free AFK Arena Diamonds' & 29 
			\\ \midrule
			7 & 'Free iTunes Gift Card' & 26
			\\ \midrule
			8 & 'Free Fortnite V-Bucks Gift Card' & 26
			\\ \midrule
			9 & 'Free Google Play Gift Card' & 26
			\\ \midrule
			10 & 'Free Starbucks Gift Card' & 25
			\\ \midrule
			11 & 'Free Roblox Robux'  & 25
			\\ \bottomrule
			\label{table:Searchterms}
		\end{tabular}
\end{table}

\paperFinding{Most scams in the dataset violated the rule on unbounded giveaways}
534 (79.9\%) of the scams in the dataset presented some form of unbounded giveaway, 
including the ``gift card generators'' which will be discussed in section~\ref{sec:case_studies}. 
54 (8.08\%) violated the rule on misrepresentation. 54 (8.08\%) others violated the rule on demonstrably invalid effects. Finally, 26 (3.89\%) others claimed to commit some sort of crime.

\subsection{RQ3: Can a classifier use metadata alone to distinguish scam and non-scam videos?}

During our follow-up analysis, we explored the feasibility of automatically distinguishing scam and non-scam videos using metadata. All analysis was done using the Python package sci-kit learn~\cite{scikit-learn}.

\begin{figure}[!tbp]
    \includegraphics[width=\columnwidth]{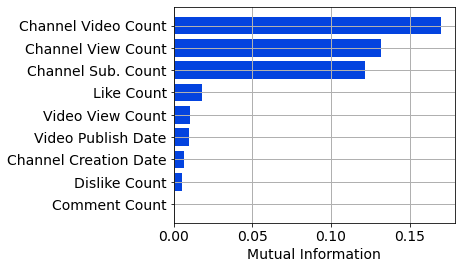}
    \caption{Mutual information estimates for the relationship between video statistics and video scam status show that statistics related to channel popularity are the most descriptive.}
    \label{fig:miuniveriate}
\end{figure}

The simplest features to analyze were the univariate numerical statistics, like those analyzed in subsection~\ref{ssec:vidstats}. We used scikit-learn's mutual information estimation to measure the dependency between it and the boolean corresponding to whether a video is a scam or not. It is equal to zero if the two are totally independent, with greater values indicating a greater dependency. The results are summarized in figure~\ref{fig:miuniveriate}. This led to the first finding:

\paperFinding{The video statistics related to channel size and popularity have the greatest predictive power} This included channel video count, channel view count, and channel subscriber count. All other features seem to provide virtually no information about whether a video is a scam, with mutual information values below 0.02. Even the peak value of 0.17 for mutual information for channel video count is not very significant. This low level of information contained within statistics was somewhat expected, as these variables are probably similar among all videos which appear high on search rankings. 

Analyzing the text-based metadata was more complicated, as text must be represented in some numerical form before analysis can be done. A word frequency or ``bag of words'' representation is the simplest, but it does not adjust for the relative significance of words; 'text' may occur many times within a document, but that high magnitude is not significant if the word appears many times in all documents. By contrast, term frequency-inverse document frequency (TF-IDF) encoding has been shown to be a robust way to ensure word magnitude is considered relative to the entire corpus of documents.~\cite{SALTON1988513}\cite{ROBERTSON2004} Each document is represented by the frequency of each word's occurrence in the document multiplied times the inverse of the magnitude of the word's occurrence across all documents. Sci-kit Learn's English stop words and the search terms were removed from all metadata before encoding. This representation was used for our analysis. 

From this point, our analysis for meta-data was very similar to the analysis of other fields. For each field, we calculated the mutual information for each TF-IDF unigram and ranked the most significant words across all videos. These results are summarized in table~\ref{table:unigram_mi}. This analysis led to our next findings:

\paperFinding{The most predictive title unigrams reflect the high prevalence of gift card and mobile game-related scams} The key words ``code(s),'' ``promo,'' ``android,'' and ``ios'' appear in the top 10 for title unigrams. Other more generic terms appear as well, including the year and words like ``hack,'' ``unlimited,'' and ``New.''

\paperFinding{The most predictive description unigrams appear to indicate that the use of links could be used as a distinguishing features} The top ten unigrams include artifacts from links, including ``com,'' ``https,'' ``www'' and ``http.'' This is slightly unusual as many of the scam videos seemed to obfuscate links or only display them in image form. 

The most predictive tag unigrams are largely unhelpful, as most tags in the dataset were related to search terms, so excluding them removed many potentially high mutual information tags. Finally, channel title unigrams are also not very useful, as particular channels which appeared many times in the dataset had an outsized influence on this experiment. 

\begin{table}[!tbp]\centering
	\caption{Unigrams which have the highest mutual information in \\ each meta-data field}\centering
		\begin{tabular}{ccccc}
			\toprule
			 & \multicolumn{4}{c}{\textbf{Field}}
			\\ \cmidrule{2-5}
			\textbf{Rank} & Title & Description & Tags & Channel Title
			\\ \midrule
			1 & hack & com & 5mp & setup
			\\ \midrule
			2 & 2019 & https & 52 & offer
			\\ \midrule
			3 & codes & video &  38aujwy & gaming
			\\ \midrule
			4 & 2020 & www & 1k & sivaji
			\\ \midrule
			5 & android & http & basics & stuff
			\\ \midrule
			6 & ios & like & appreciated & code
			\\ \midrule
			7 & unlimited & channel & asksebby & new
			\\ \midrule
			8 & new & subscribe & capital & wired
			\\ \midrule
			9 & code & youtube & channels & today
			\\ \midrule
			10 & promo & follow & appbounty & game
			\\ \bottomrule
			\label{table:unigram_mi}
		\end{tabular}
\end{table}

\begin{figure}[!tbp]
    \includegraphics[width=\columnwidth]{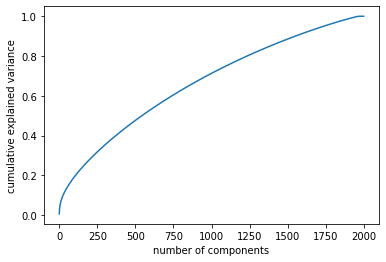}
    \caption{This graph shows the cumulative explained variance per number of components for captions after SVD. It shows that significant dimensionality reduction is not possible without loosing a large amount of descriptive power}
    \label{fig:svd}
\end{figure}

English captions were present on roughly 2/3 of the remaining videos in the dataset. Their prevalence was roughly equal between scam and non-scam videos. Most captions were automatically generated and, therefore, not extremely reliable. We attempted dimensionality reduction on the captions using singular value decomposition (SVD) for the sparse matrix of captions, but it was not possible to reduce the data efficiently. As shown in figure~\ref{fig:svd}, over 1700 components were still required to explain 90\% of the variance in the captions between scam and non-scam videos. We determined that this reduction was not large enough to make it possible to analyze captions further.

The final part of our dataset which has not been discussed is video information, including the video file and the audio captions. Unfortunately, due to the sparsity of our dataset, we did not have enough data to explore classification on a high-dimensional space like video. 

Leaving video analysis for future work, we looked for simple heuristics which may help with the classification task. Video and audio statistics like bit rate, frame rate, creation date, and resolution did not seem to contain any useful information. Ultimately, this simple analysis of video data did not prove it to be useful for classification, but more may be found with a larger dataset or if the more accurate transcription is used in future work.
\section{Case Studies}
\label{sec:case_studies} 
The first case study described 
in subsection \ref{ssec:generator} is a distinct type of 
fraud which we have termed ``gift card generators.'' The second case study relates to tech support scams that use misrepresentation to fool users into calling a phone number or visiting a website. These are discussed in 
subsection \ref{ssec:techsupport}. The final case study discussed in subsection \ref{ssec:fortnite} involves the clickbait scam
videos created by several moderately large YouTube channels.

\subsection{Gift Card Generators}
\label{ssec:generator}
\paperFinding{Of the 668 scam videos in our dataset, 414 (62.0\%) of them redirected to websites which claim to ``generate'' gift cards or mobile game
resources, usually by ``hacking''} 
We have termed these ``gift card generators.''
Before these gift card generators provide the currency
or gift cards that they claim they can produce, they state that
``human verification'' is required to prevent overuse of the tool. 
This ``verification'' is achieved through a web survey or application download, likely providing a stream of income to scammers. The videos try to increase the realism of these scams by
showing a gift card being redeemed or in-game currency rapidly increasing.

Gift card generators often pretend to try automatic verification before requiring a 
survey or app download. In this way, these scams take advantage
of fake feedback to try to fool users into thinking the scam is legitimate. While this particular type of scam does not seem to harm a user's
device or extract private information, it does mislead users and uses them to earn money. Additionally, with the scam's prevalence
on terms related to mobile games, young adults and even children may be particularly vulnerable.

\begin{figure}[!tbp]
    \includegraphics[width=\columnwidth]{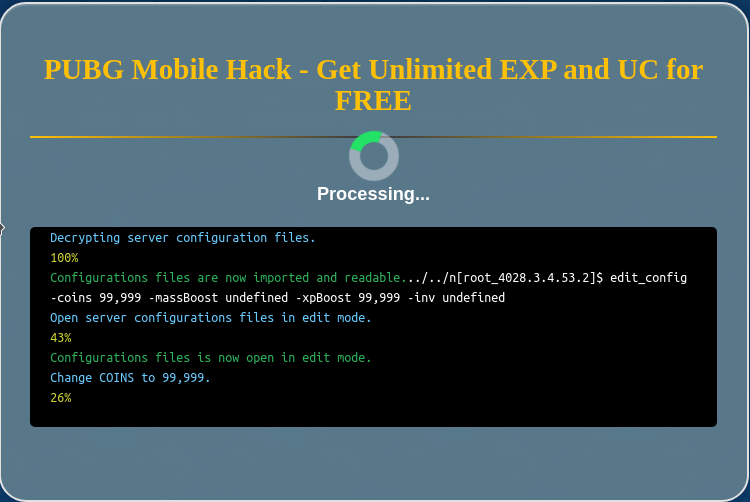}
    \caption{Example of a gift card generator animation which was seen on several websites. The graphic shows a loading bar and simulated console output to increase its perceived validity. One the ``loading'' completed, the website prompted the user to complete a survey or task.}
    \label{ggenerator_example}
\end{figure}

\paperFinding{Some scams may be facilitated by crimeware vendors} Many of the scam landing pages share similarities or even identical graphics (see an example in figure~\ref{ggenerator_example}). We traced these back to a forum centered around creating websites for ``cost-per-action'' monetization or CPA.\footnote{https://www.cpaelites.com/} While CPA can be a legal business practice employed to avoid risk and improve return on investment for advertisers~\cite{Nazerzadehetal}, those in the CPA communities behind these gift card generators employ unethical or illegal tactics to convince users to complete actions---in this case, the ``human verification'' task---and gain profit. For these gift card generator websites, YouTube is another vector for spreading the scam, and discussion on the aforementioned community includes buying YouTube views and interaction to boost return on investment.

\subsection{Tech Support Scams}
\label{ssec:techsupport}

\paperFinding{We found very few videos falsely claiming to be a legitimate company's tech support, and many of those that did exist were on just two search terms} While 33 of the 74 search terms used to generate 
our dataset were related to tech support, only 8 
of these search terms contained a scam video.
65\% of the scams were on just two tech support search terms:
``Office.com Tech Support'' and ``Yahoo.com Tech Support.''
Other search terms related to tech support were filled 
with legitimate tech support advice or videos completely
unrelated to tech support. Tech support search queries 
related to social networking sites like ``Facebook.com Tech Support'' 
or ``Twitter.com Tech Support'' were filled with unrelated videos
due to the search engine presenting videos with metadata linking
to social media accounts.

The scams that do exist appear to be tech support scams
like those discussed by Miramirkhani et al.~\cite{dial_one} While it 
is possible that some of the providers listed in these 
videos are legitimate, they cross the ethical line by fraudulently pretending 
to be a company which they are not. Indeed, there 
were several videos in the dataset which were clearly third party tech support, and therefore
were not considered scams. 

It is not immediately clear why most of the tech support scams were on only two search terms. It is noted that 
Yahoo does not have a support number that you can call, nor 
do they have a way to request that support calls you. It is 
possible that this gap allows scammers to fill the void that 
might otherwise be filled by legitimate support. 

\subsection{Clickbait Scams}
\label{ssec:fortnite}
\paperFinding{There were several cases of large YouTube creators which used clickbait to scam their audience}
While viewing videos under the search term ``Free Fortnite V-Bucks Gift Card,'' several highly similar scams were found. These videos all had titles suggesting they could teach the viewer how to generate V-Bucks---the in-game currency of the game Fortnite. 
Most of the video would continue the ruse, with the narration describing how the viewer needs to watch until the end to learn the trick. Many of these videos would instruct the viewer to add a ``Creator Code'' to their Fortnite game in order to execute the trick,
referring to the Epic Games ``Support-a-Creator Program''\footnote{\url{https://www.epicgames.com/affiliate/en-US/overview}} which provides money to online content creators when their viewers spend V-bucks. The end of the video would reveal that the title is misleading, either showing a fake glitch that fails to do anything or suggesting that players can sell promotional items to earn V-bucks.

While this may be considered simply benign clickbait, the suggestion 
that paying the creator is necessary pushes this over the edge into fraud. This scam is particularly insidious when one considers that 1) the game is frequently played by children and young teens, and 2) the creators hosting these scams have a large fan base and may be trusted. Since the initial dataset creation, these particular videos are still hosted on YouTube.
\section{Discussion and Future Work}
\label{sec:conclusion}

While to the best of our knowledge, we are the first to analyze this question in the research literature,
we surely expect that engineers at YouTube are aware of the problems of scams on their platform.
While YouTube provides no details on scam detection and removal, we suspect the high number of videos taken down between our two data collection periods may at least partially reflect an automated or manual detection process.
We have also found a discussion of evading YouTube's anti-fraud system on CPA forums.
As a consequence, the scams that we analyze are likely 
``survivors'' of whatever processes are in place.
It may be the case that the scams we study remained because they succeeded
in maintaining a reasonable like/dislike ratio, making their metadata similar to legitimate content, or because they had not yet met some popularity threshold that would trigger a review. Our finding that metadata alone is unlikely to identify scams may only apply to these videos because videos that \emph{could} be detected easily have already been removed.
Our limited sample also focused primarily on topics likely to have associated fraud in order to acquire a large enough number of scam videos.
Because of this selection bias and survivor bias, we are likely seeing only a narrow window into a larger phenomenon.

 Future work should look at collecting larger datasets and  using additional features beyond those analyzed in this work, including 
 multimedia video and audio analysis.
 Additionally, future work could look more deeply into the world of fraudulent CPA practices, as many of the scams on YouTube are based upon those revenue streams. It may be possible to focus on identifying the crimeware used for these scams to better detect them automatically.
Future work could also look more deeply at potential connections between campaigns to more accurately capture the ecosystem.  Finally,
we observe that individuals, consumer advocates, and regulators would benefit from greater transparency from YouTube and other video platforms about the quantity and tactics of scams on their platforms.

\section*{Acknowledgements}

The authors thank our anonymous reviewers for their helpful insights. 
This material is based upon work supported by the National Science Foundation under Grant No. CNS-1849994. 

\bibliographystyle{IEEEtran}  
\bibliography{main.bib}
\end{document}